\documentclass[twocolumn,showpacs,preprintnumbers,amsmath,amssymb,floatfix,showkeys]{revtex4}
\usepackage{amsfonts}
\usepackage{txfonts}

\usepackage{graphicx}
\usepackage{dcolumn}
\usepackage{bm}
\usepackage[center]{subfigure}

\begin{document}


\title{Atomic decoration for improving the efficiency of field electron
emission of carbon nanotubes}

\author{Guihua Chen }
\author{Zhibing Li }
 \email{stslzb@zsu.edu.cn }
\author{Jie Peng, Chunshan He, Weiliang Wang,\\ Shaozhi Deng }
\author{Ningsheng Xu }
 \email{stsxns@zsu.edu.cn }
\affiliation{State Key Laboratory of Optoelectronic Materials and Technologies School of Physics and Engineering, Sun
Yat-Sen University, Guangzhou 510275, China }

\author{Chongyu Wang and Shanying Wang }
\affiliation{Department of Physics, Tsinghua University, Beijing 100084, China }

\author{Xiao Zheng and GuanHua Chen }
\affiliation{Department of Chemistry, the University of Hong Kong, Hong Kong, China }

\author{Tao Yu }
\affiliation{Central Iron and Steel Research Institute, Beijing, China }

\date{\today} 

\begin{abstract} The field electron emission from the single-walled
carbon nanotubes with their open ends terminated by -BH, -NH, and -O has been simulated. We find that -BH and -NH
suppress the apex-vacuum barrier significantly and lead to higher emission current in contrast to the -O terminated
structure. The calculated binding energy implies that the carbon nanotubes terminated with -BH and -NH are more stable
than those saturated by oxygen atoms or by hydrogen atoms. The simulation shows that the most probable orientation of
the emission beam has correlation to the atomic structure of the apex and is field-dependent.
\end{abstract}

\keywords{AVB, adsorption, electronegativity, emission path}
\pacs{73.22.-f, 73.21.-b, 73.63.Fg, 79.70.+q} \maketitle

\section{\label{sec:level}Introduction}

The field electron emission (FE) from carbon nanotubes (CNTs) has found its applications in flat panel displays
\cite{a1} and in molecular sensors. \cite{a2,a3,a4} It has also shown potential applications in miniature
highbrightness electron sources for electron microscope \cite{a5} and in parallel e-beam lithography system. \cite{a6}
One of the central problems in these applications is to optimize the structure to achieve stronger FE in lower
macroscopic field. It was once believed that the large aspect ratio of a CNT that would lead to large local field
enhancement (LFE) at the apex of the CNT was the major reason responsible for the superior FE, hence it would be
straight forward to improve the FE efficiency by increasing the length of the CNT. But this common view has not been
verified satisfactorily by either experimental or theoretical studies. Recent simulations have shown that the charge
accumulation in both the apex and the body of the CNT can significantly modify the apex-vacuum barrier (AVB) and
thereby the FE characteristics. \cite{a7,a8} A recent simulation found that the field enhancement factor for the open
single-walled CNT (SWCNT) is much smaller than the expected value of the metal rod model; on the other hand, the change
of the AVB has more pronounced effect to FE. \cite{a9}

Since a CNT emitter has only decades of atoms at the apex, the atomic structure of the apex would have strong influence
on the FE process. The following effects have been observed in experiments: the hydrogenation of the tube wall
transforms a metallic CNT to a semiconductive one, \cite{a10} O$_2$ exposure increases the turn-on field of SWCNTs and
decreases the FE efficiency, \cite{a11} and the adsorption of H$_2$O enhances the field emission current. \cite{a12}
However, experimental observations for the adsorbate effects so far have not been conclusive. \cite{a11,a13,a15} To
understand the dependence of FE upon the atomic structure of the apex, careful simulations via the density functional
method (DFT) have been carried out. \cite{a16,a17,a18,a19,a20,a21,a22} However, there are contradictory conclusions
about the effect of adsorbates. Zhou \textit{et al.} \cite{a23} and Kim \textit{et al.} \cite{a7,a24} obtained the
local density of states (LDOS) at the apex by the ab initio methods and found that the LDOS at the charge-neutrality
level is suppressed by the hydrogen. They therefore concluded that hydrogen adsorption reduces the FE current density.
By contrast, Mayer \textit{et al.} calculated the AVB using a dipole and point charge model. \cite{a7} They acclaimed
that the apex-vacuum barrier is reduced by the presence of the hydrogen, and thereby concluded that hydrogen adsorption
would enhance the FE current density. Mayer recently improved the model and illustrated the electrostatic potential
around the carbon nanotube. \cite{a10} More careful studies on this topic would obviously be useful.

Only recently, it is possible to tackle the SWCNT of realistic size in the FE conditions by a multi-scale method
involving quantum mechanics and molecular mechanics. \cite{a8,a26} In the present paper, we have adopted this method to
simulate the FE from the SWCNT of realistic length (one micrometer) with different atomic decorations at the open end
of the tube. Obviously the AVB and thereby the FE characteristic will be strongly affected by the electron transfer
between the carbon atoms and the adsorbent atoms. The atomic decoration in the apex will induce dipoles due to the
geometric symmetry breaking in the axial direction. If the dipole has its positive end outward to the vacuum (positive
dipole), it tends to suppress the AVB; otherwise (negative dipole), it inclines to raise the barrier. This simple
argument suggests that the carbon dangling bonds in the open end of the SWCNT should be saturated by atom with
electronegativity ($\mathit{X_S}$) lower than the carbon atom. For instance, the hydrogen terminated SWCNT has a
positive dipole because hydrogen has lower $\mathit{X_S}$ than carbon. \cite{a9} The electronegativities of hydrogen,
boron, carbon, nitrogen, and oxygen in Allen electronegativity scale are given in TABLE \ref{tab_I}.

At room temperature, the electrons in the vicinity of the apex are most relevant to the emission. When the apex has
higher density of electrons, there are more incident electrons hitting on the AVB and a stronger emission current could
be expected. To attract electrons from the tube column to the apex, it would be useful to saturate the CNT first by
atoms of higher $\mathit{X_S}$ then followed by atoms of lower $\mathit{X_S}$. Another consideration is the structural
stability that is one of the most concerned properties in the applications. In this paper, we should consider the
diatom ions -BH and -NH as adsorbates and compare them with the oxygen ion. Our simulation should show that the
adsorbates of -BH and -NH  have larger binding energy than both oxygen and hydrogen.

\begin{table}
\caption{\label{tab_I} Electronegativity of related atoms in Allen electronegativity scale.}
\begin{ruledtabular}
\begin{tabular}{cccccc}
atom\ &H\ &B\ &C\ &N\ &O\\
\hline
$\mathit{X_S}$ & 2.300 & 2.051 & 2.544 & 3.066 & 3.610\\
\end{tabular}
\end{ruledtabular}
\end{table}

We should adopt the multi-scale method \cite{a8,a26} to simulate the (5, 5) SWCNT with its open end saturated  in three
kind of ions: -BH, or -NH, or -O. In Section II, the method of the simulation is reviewed briefly. The charge density
and electrostatic potential in the absence of applied field is given in Section III. In Section IV, the field-depending
AVB is illustrated. The most probable emission path is studied and the emission characteristics of the (5, 5) SWCNTs
saturated with -BH, -NH and -O are presented in Section V. The last section is devoted to the discussions and
conclusions.

\section{\label{sec:level}Simulation method and the ending structures}

\begin{figure}
\includegraphics[scale=0.8]{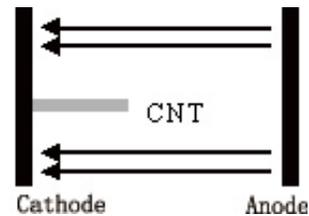}
\caption{\label{fig_1} The schematic setup of field electron emission. The arrow lines represent the applied electric
field.}
\end{figure}

The concept of multi-scale coupling is important for the simulation of huge systems that are sensitive to all scales of
the systems. \cite{a27,a28} The CNTs for the purpose of FE are typical multi-scale systems. A simplified schematic
setup of the FE system is shown in FIG. \ref{fig_1}, in which two black plates are the cathode (left) and anode
(right); a CNT is mounted vertically on the cathode. When a voltage is applied to the two metal plates, the electrons
have opportunity to emit into vacuum through the apex of the CNT by quantum tunneling.

In experiments, the length of CNTs is usually in micrometers, while the radius is in nanometers. For instance, the (5,
5) type SWCNT of one micrometer long consists of about \textit{10000} carbon atoms. Limited by the computational
efficiencies and resources, all ab initio studies so far can only simulate the local properties involving hundreds of
carbon atoms. As the electronic properties are sensitive to both the detailed atomic arrangement (i.e., the location of
defects and adsorbates) and the electron states, which would extend through the whole tube, it is a big challenge to
simulate the whole tube that has a length in micrometer scale.

Since electrons are emitted from the apex of the SWCNT by quantum tunneling, the apex part must be treated by quantum
mechanics. The part on the substrate side mainly affects the FE through Coulomb potential of the excess charge, so that
it can be treated by a semi-classical method. \cite{a8,a26} Therefore we divide the SWCNT into a quantum region and a
semi-classical region. The quantum region is simulated on atomic scale where the density matrix of electron is obtained
quantum mechanically. In the semi-classical region, the Coulomb potential is governed by the Poisson's equation.

It should be noted that even in the semi-classical region the electron energy band structure originated from quantum
mechanics should be taken into account. The excess charge distribution in the semi-classical region has simple solution
only for the simple band structure. For the (5, 5) SWCNT, there are both experimental and theoretical evidences for the
constant density of states (DOS) in the vicinity of neutrality level. \cite{a29,a29_I,a29_II,a30} Hence the
semi-classical region should be limited in the region where the constant DOS is valid. To our experience, this means
that the length of the quantum region should be over \textit{100}\textrm{ nm}.

In the present simulation, the quantum region extends \textit{123}\textrm{ nm} and contains \textit{10000} atoms. It is
still too big for an ab initio simulation. The quantum region is then further divided into sub-regions. Each sub-region
together with its adjacent sub-regions as buffer form a subsystem that is simulated by the modified neglect of diatomic
overlap (MNDO) semi-empirical quantum mechanical method (here the MOPAC software had been used). The excess charges in
the quantum region but not in the subsystem being simulated are treated as point charges. Their contribution to the
subsystem being simulated is through Coulomb interaction.

The coupling of the quantum region and the semi-classical region is through the quasi-thermodynamic equilibrium
condition which assumes that the electrochemical potential (Fermi level) is a constant along the nanotube, and equal to
that of the substrate. The densities of the excess charge ("excess density" for simplicity) calculated separately in
quantum and semi-classical regions should coincide at some overlap place of two regions. Here we required that the
excess densities coincide at the position of \textit{900}\textrm{ nm} from the substrate. The self-consistent excess
density of the entire CNT is achieved through iterations that contain a small loop and a big loop. In the small loop,
the sub-regions of the quantum region are simulated one by one, and repeated until that a converged electron density of
the quantum region is obtained. In the big loop the quantum region and the semi-classical region are simulated
alternatively until the self-consistent charge distribution is achieved. This process is greatly accelerated by the
observation that the charge density in the region far from the apex, to a good approximation, is a linear function of
the distance from the substrate. This linear behavior is a consequence of the constant DOS in the classical region.
\cite{a31}

When the SWCNT is mounted on the metallic substrate, Schottky junction will be formed at the back contact in principle.
However, as the SWCNT here is very long, the effect of this junction could be ignored. Equivalently, we have assumed
that the Fermi level of the isolated SWCNT aligns with that of the metal. The boundary condition of the metal surface
is guaranteed by the image charges of the excess charges of the SWCNT. The total electrostatic potential is the
superposition of the Coulomb potential created by excess charges and their images in the substrate, as well as the
applied macroscopic field.

\begin{figure} 
\centering \subfigure[]{
\label{fig_2_a} 
\includegraphics[scale=0.165]{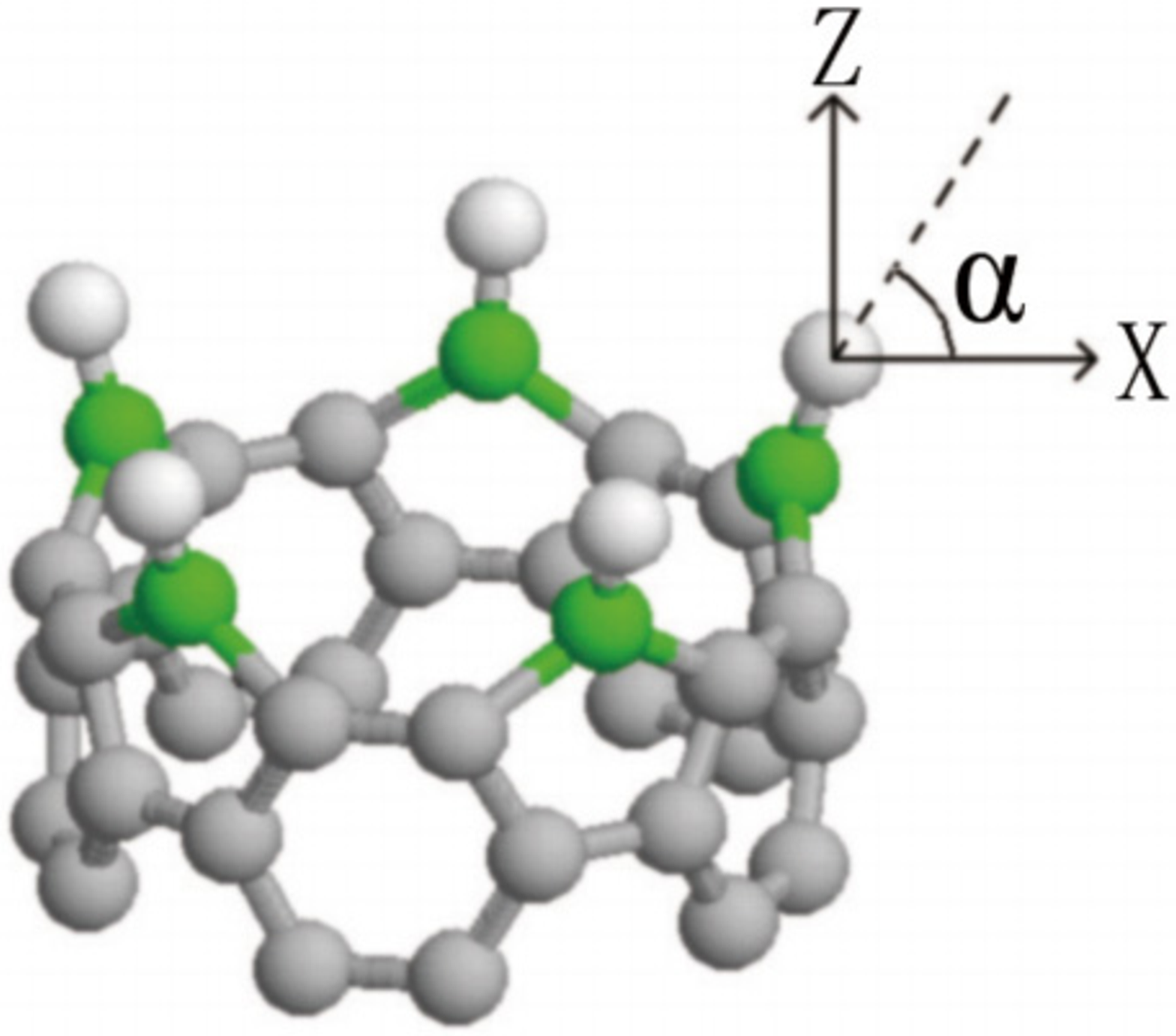}}
\hspace{0.02in} \subfigure[]{
\label{fig_2_b} 
\includegraphics[scale=0.16]{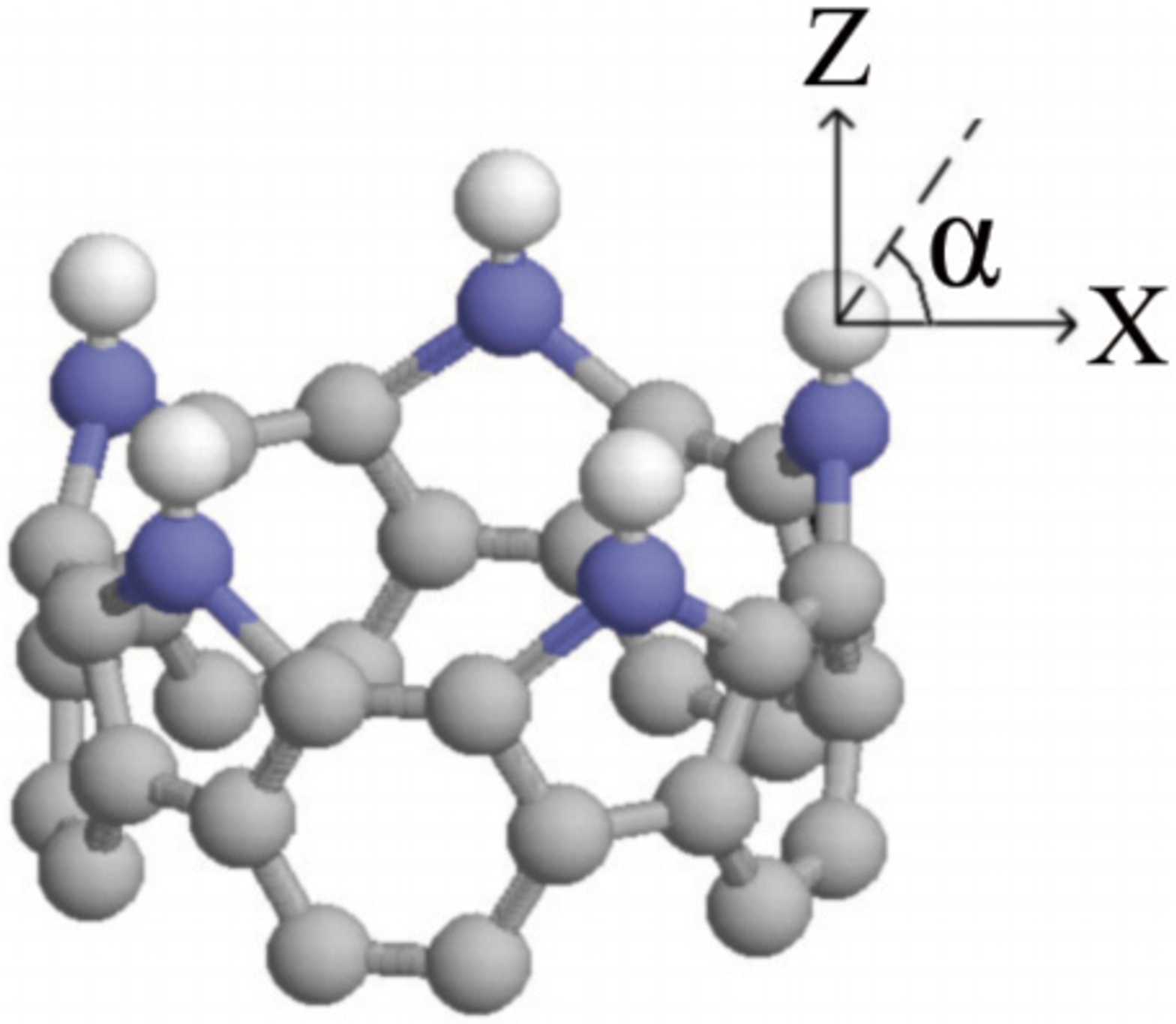}}
\hspace{0.02in} \subfigure[]{
\label{fig_2_c} 
\includegraphics[scale=0.16]{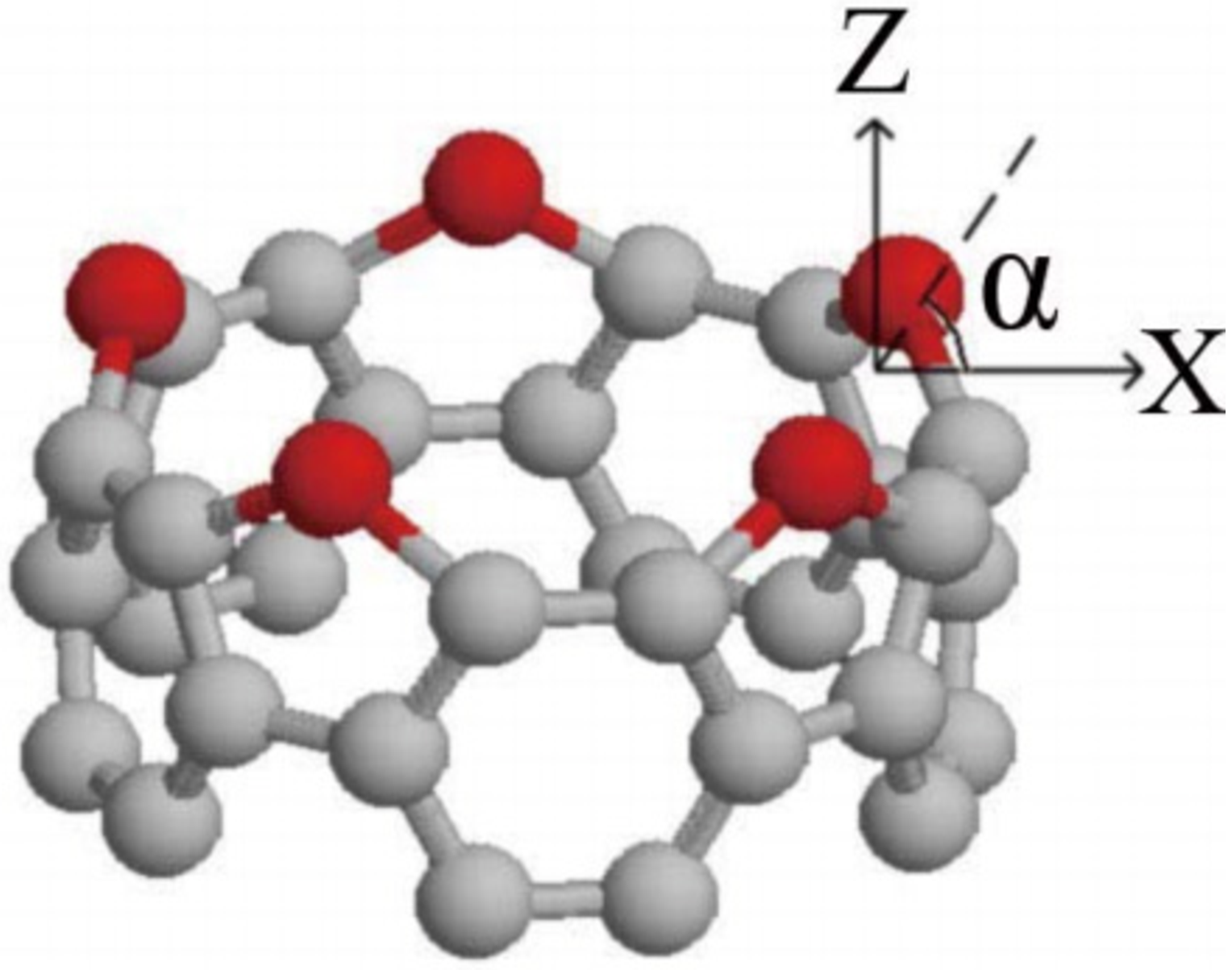}}
\caption{Three ending structures of the (5, 5) SWCNT. The gray balls stand for carbon atoms. (a) The a-type structure
with -BH as adsorbates. The green balls stand for boron atoms and white balls for hydrogen atoms. (b) The b-type
structure with -NH as adsorbates. The blue balls and the white balls denote nitrogen atoms and hydrogen atoms,
respectively. (c) The c-type structure with oxygen ions as adsorbates. The red balls stand for oxygen atoms. The
\textit{Z}-axis in each figure is parallel to the axis of the tube. The \textit{X}-axis is the radial axis on the plane
perpendicular to the tube axis and through a hydrogen atom (for FIG. \ref{fig_2_a} and \ref{fig_2_b}) or through the
middle point of two adjacent oxygen atoms (for FIG. \ref{fig_2_c}). The angle $\alpha$ is on the \textit{X}-\textit{Z}
plane.} \label{fig_2}
\end{figure}

The apex structures with -BH, -NH, and -O terminations are shown in FIG. \ref{fig_2}; they are referred to as a-type,
b-type, and c-type structures respectively. The five green balls in FIG. \ref{fig_2_a} for the a-type structure stand
for the boron atoms. Each boron atom shares two bonds with two carbon atoms. The third bond of boron atom is saturated
by a hydrogen atom (the white ball). Similarly, the b-type structure is shown in FIG. \ref{fig_2_b}, with the blue
balls for nitrogen atoms and white balls for hydrogen atoms. The five oxygen atoms are represented by red balls in FIG.
\ref{fig_2_c} for the c-type structure. All ending structures are relaxed with the MOPAC software. The coordinates of
adsorbent atoms and the first layer of carbons are given in TABLE \ref{tab_II}.

\begin{table}
\caption{\label{tab_II} The coordinates of the adsorbates and the carbons of the top layer. Where \textit{Z} stands for
the axial coordinate with the origin on the substrate surface, \textit{X} and \textit{Y} denote two orthogonal
coordinates of the plane perpendicular to the tube axis, with the origin at the axis.}
\newcommand{\rb}[1]{\raisebox{-2.5ex}[0pt]{#1}}
\newcommand{\rc}[1]{\raisebox{1.0ex}[0pt]{#1}}
\begin{ruledtabular}
\begin{tabular}{lcccc}
\multicolumn{2}{c}{\rb{Structures/atoms}}&\multicolumn{3}{c}{Coordinate (angstrom)}\\
\cline{3-5} & & Z & X & Y\\
\hline & H & 10088.04 & 2.35 & 2.85\\
-BH & B & 10086.92 & 2.10 & 2.53\\
& C & 10086.04 & 0.94 & 3.31\\
\hline & H & 10087.60 & 2.35 & 2.83\\
-NH & N & 10086.87 & 1.91 &  2.31\\
& C & 10086.04 & 0.94 & 3.31\\
\hline  & O & 10086.83 & 2.05 & 2.46\\
\rc{-O} & C & 10086.04 & 0.94 & 3.31\\
\end{tabular}
\end{ruledtabular}
\end{table}

The binding energy $E_{bind}$ is defined as

\begin{equation}\label{eq_1}
E_{bind}=E_{sep}-E_{bond},
\end{equation}

\noindent where $E_{sep}$ is the total energy of the system in which the adsorbates are separated far away from each
other and from the CNT, and $E_{bond}$ is the total energy of the system in which the adsorbates are bonding with the
carbons of the top layer. In the calculation of binding energies, we only consider the last ten layers of carbons at
the tip. The stability of the ending structure is described by the binding energy per adsorbent diatom of the
a-type/b-type structure or per adsorbent atom for the c-type structure. The results are given in TABLE \ref{tab_III}.
The a-type and b-type ending structures are more stable than the c-type structure and the structure in which every
dangling bond of carbon is saturated by a hydrogen atom.

\begin{table}
\caption{\label{tab_III} The binding energy of adsorbates.}
\begin{ruledtabular}
\begin{tabular}{ccccc}
ions\ &-BH\ &-NH\ &-O\ &-H\\
\hline binding energy (eV) & 17.8 & 18.3 & 3.99 & 8.63\\
\end{tabular}
\end{ruledtabular}
\end{table}

\section{\label{sec:level}Electronic structures in the absence of applied field}

The FE characteristic of each individual SWCNT should be relied on its intrinsic electronic structure. Therefore in
this section we focus on the electronic structures of the SWCNT in the absence of external macroscopic field.

\begin{figure} 
\centering \subfigure[]{
\label{fig_3_a} 
\includegraphics[scale=0.38]{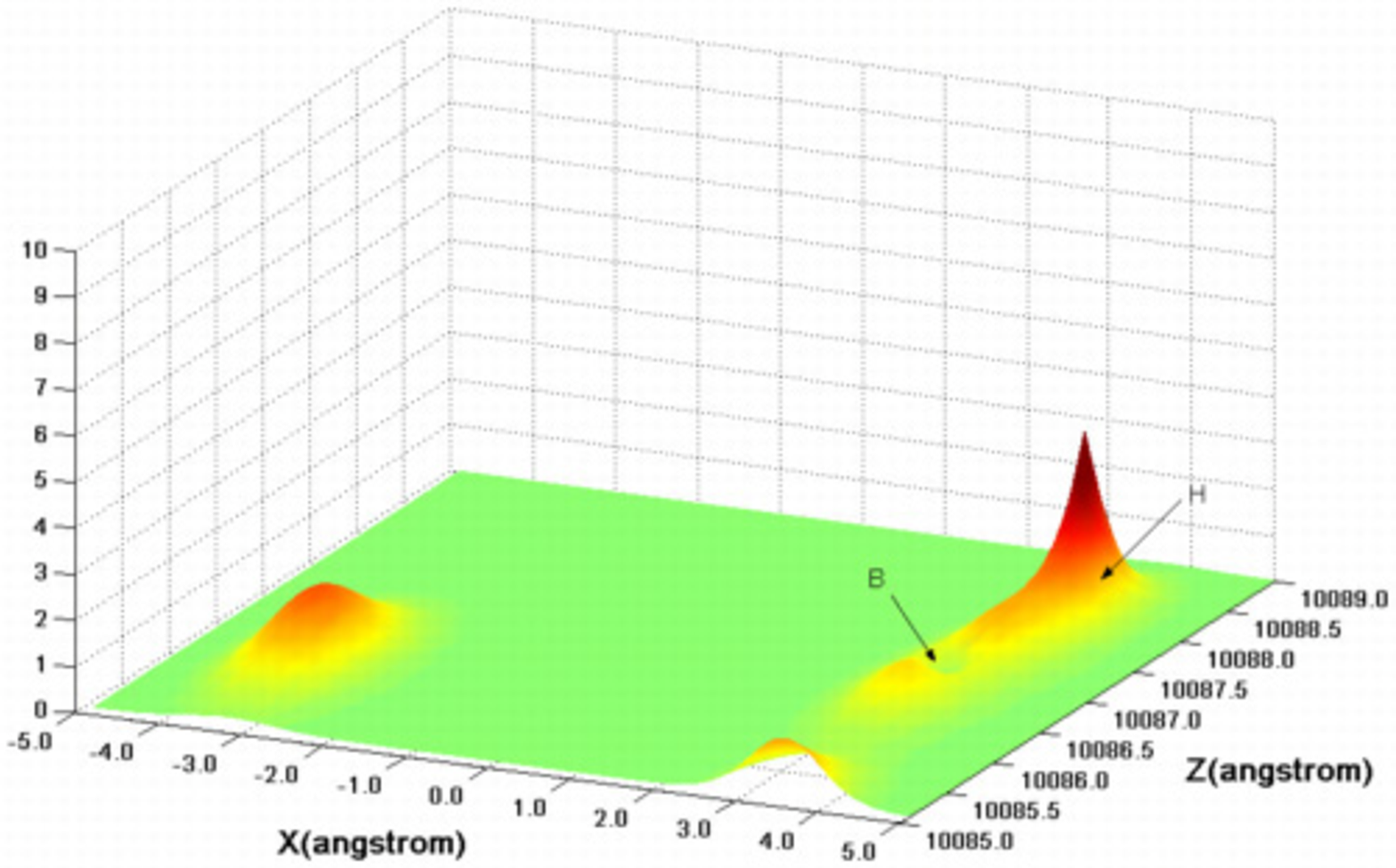}}
\hspace{0.1in} \subfigure[]{
\label{fig_3_b} 
\includegraphics[scale=0.38]{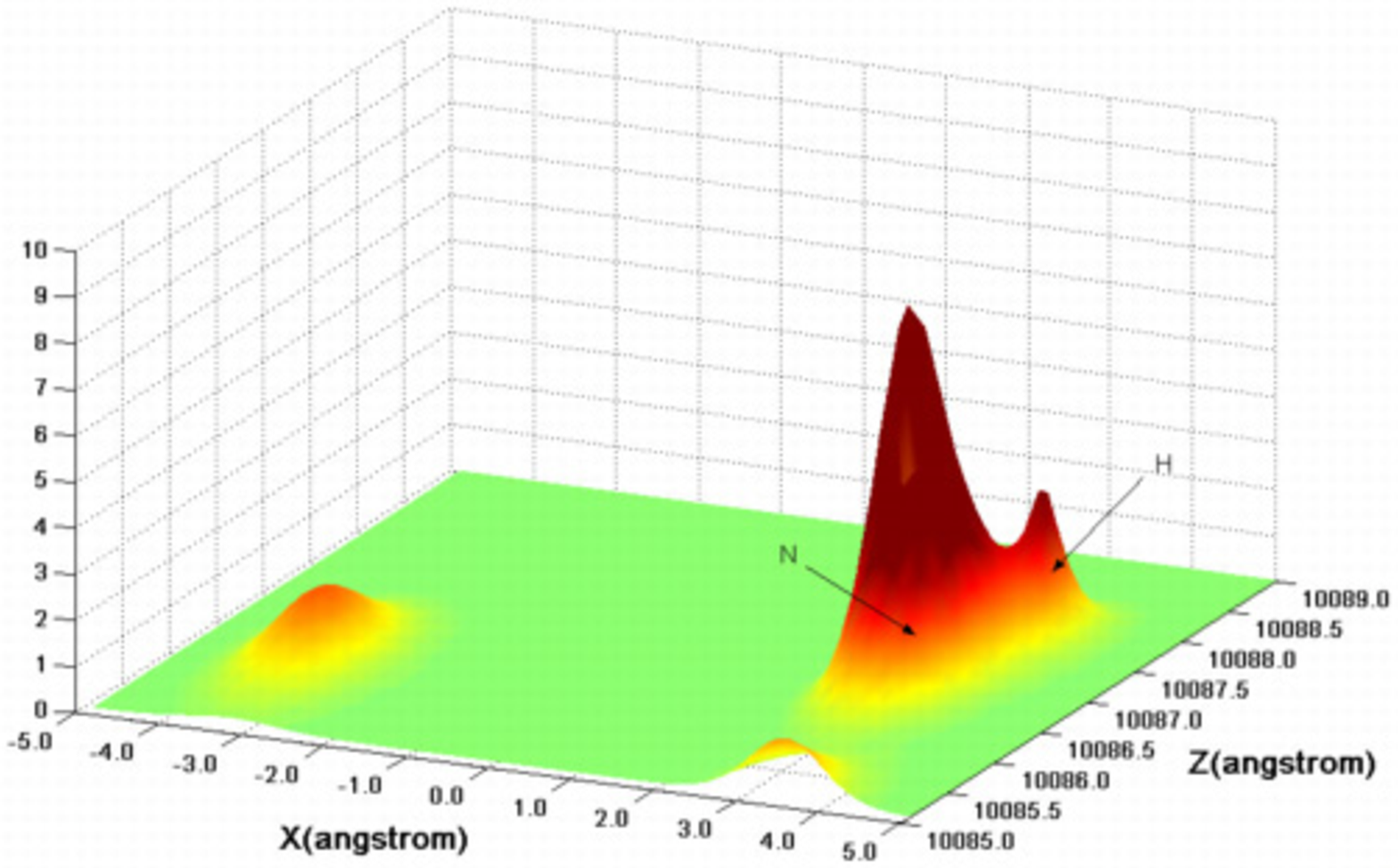}}
\hspace{0.1in} \subfigure[]{
\label{fig_3_c} 
\includegraphics[scale=0.38]{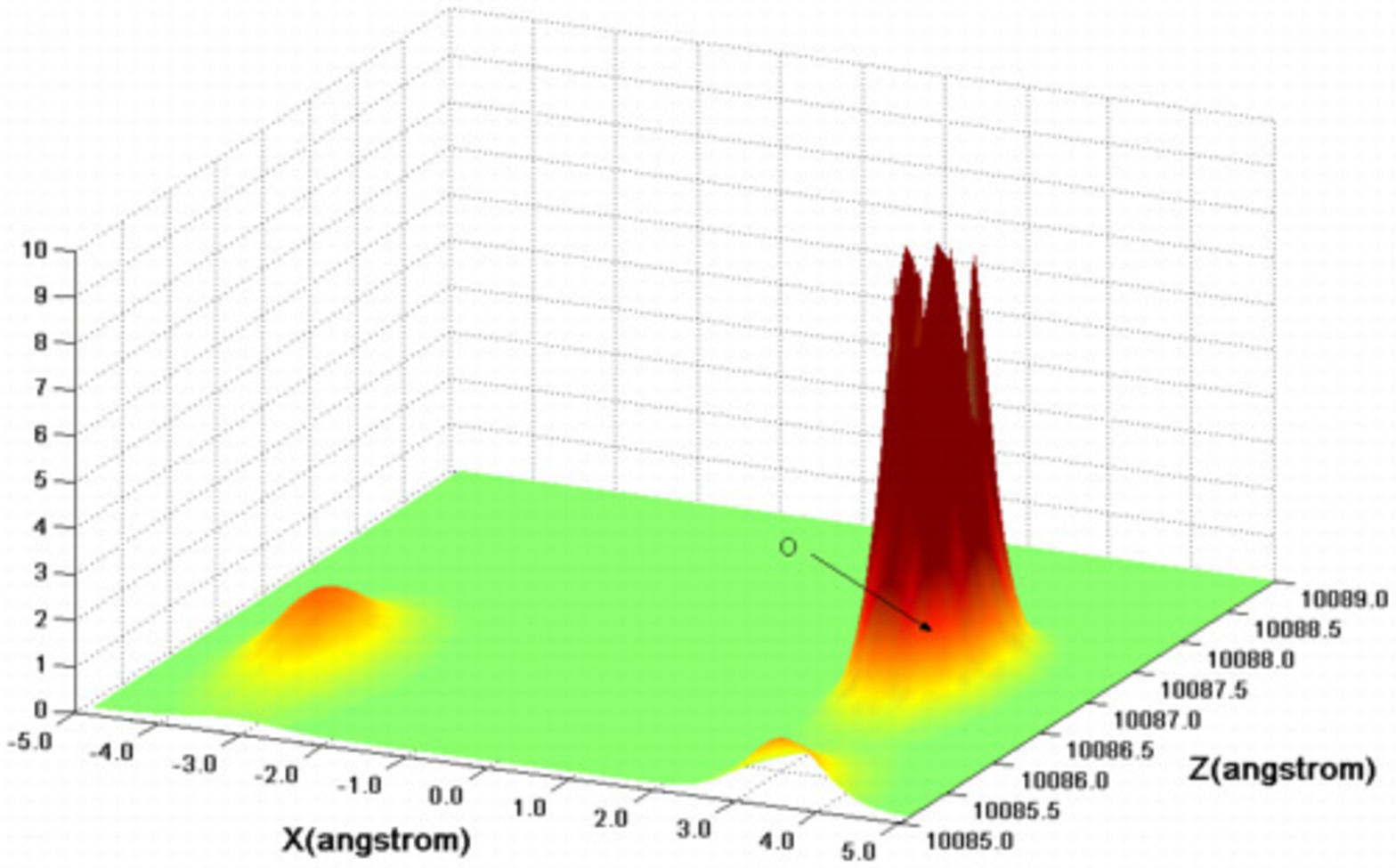}}
\caption{The distribution of electrons on the intersection plane crossing one adsorbate atom and the axis of the SWCNT.
FIG. \ref{fig_3_a}, \ref{fig_3_b}, and \ref{fig_3_c} are corresponding to the a-type, b-type, and c-type structures
respectively.}
\label{fig_3} 
\end{figure}

\begin{figure}
\includegraphics[scale=0.36]{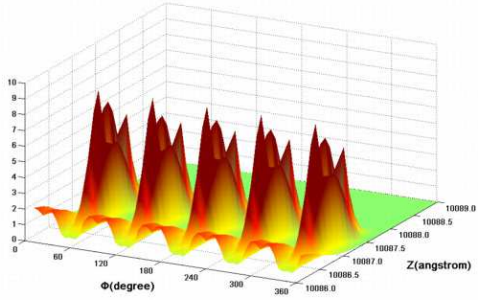}
\caption{\label{fig_4} The electron distribution around the atoms of last two layers of the SWCNT terminated with
oxygen atoms.}
\end{figure}

FIG. 3(a)/(b)/(c) shows the electron density in the intersection plane, on which the axis of SWCNT and one
boron/nitrogen/oxygen atom are located. The arrows mark the positions of the adsorbent atoms. In the a-type structure
(FIG. \ref{fig_3_a}), electrons obviously transfer from the boron atoms to the hydrogen atoms. From FIG. \ref{fig_3_b},
one can see that electrons in the b-type structure are concentrated at the nitrogen atoms. The electrons are strongly
concentrated at the oxygen atoms in the c-type structure (FIG. \ref{fig_3_c}). The electron distribution in the c-type
structure can be seen more clearly in FIG. \ref{fig_4}, where the distribution of excess electrons is shown along the
wall of the SWCNT. The excess charge associated to each atom of the top three(two) layers in the a-type/b-type(c-type)
structure has been calculated by the Mulliken population analysis method. The results are given in TABLE \ref{tab_IV}.
Since the boron has smaller electronegativity in comparing with the carbon and the hydrogen, electrons tend to transfer
from the boron to both the carbon and the hydrogen in the a-type structure and lead to a dipole with its negative
charge outward. In the c-type structure, we find a dipole of the same direction as the dipole in the a-type structure
with a bigger value. This is expected as $\mathit{X_S}$ of the oxygen is much larger than that of the carbon. A
reversed dipole, i.e., with its positive charge outward, arises in the b-type structure due to the large
electronegativity of the nitrogen. The dipole values of the top three (two) layers of the a-type and b-type (c-type)
structures are given in TABLE \ref{tab_V}.

\begin{table}
\caption{\label{tab_IV}The distribution of the excess charges.}
\begin{ruledtabular}
\begin{tabular}{ll}
atom group & excess charges (e) \\
\hline CBH & -0.082(C), 0.276(B), -0.039(H) \\
CNH &  0.086(C),-0.248(N), 0.150(H) \\
CO &  0.182(C),-0.302(O)\\
\end{tabular}
\end{ruledtabular}
\end{table}

\begin{table}
\caption{\label{tab_V} The dipoles in the first and the second top layers.}
\newcommand{\rb}[1]{\raisebox{-2.0ex}[0pt]{#1}}
\newcommand{\rc}[1]{\raisebox{1.0ex}[0pt]{#1}}
\begin{ruledtabular}
\begin{tabular}{ccc}
\multicolumn{1}{c}{\rb{Structures}} &\multicolumn{2}{c}{dipole (e\textperiodcentered nm)} \\
\cline{2-3} & atom of 1st layer & atom of 2nd layer \\
\hline -BH & -0.0095 &  0.0135 \\
-NH &  0.0158 & -0.0122 \\
-O &  -0.0154 & -0.0051 \\
\end{tabular}
\end{ruledtabular}
\end{table}

\section{\label{sec:level}Field-depending barrier}

Now let us investigate the response of the SWCNT to the macroscopic fields (\textit{F}$_{appl}$). The general picture
is as follows. The applied field \textit{F}$_{appl}$ drives electrons to the tip of the SWCNT until the equilibrium is
set up. We assume that the emission current is weak and can be ignored in the calculation of the electron density of
the SWCNT. The redistribution of electrons in \textit{F}$_{appl}$ has two consequences. First, the field is more or
less shielded in the body of the tube. This leads to the field enhancement at the apex. However, as the wall of the
SWCNT has only one layer of atoms, the shielding would be not complete, especially in the apex region where the field
is strong. There would be a field penetration in the apex region, so the field enhancement factor is smaller than that
predicted by the classical theory for a metal rod. Second, there are excess charges accumulating along the tube,
especially in the apex region. In the quasi-equilibrium assumption, the charge accumulation is possible only if the
neutrality energy level of the SWCNT bends down. This happens when the applied field lowers the energies of local
orbitals so that the orbitals with their energies lying between the Fermi level of the substrate and the neutrality
energy level can accommodate electrons and contribute to excess charges.

\begin{figure}
\includegraphics[scale=0.7]{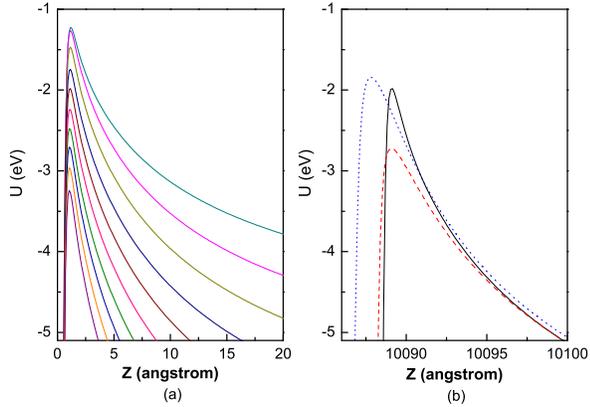}
\caption{\label{fig_5} The AVB of the SWCNT. (a) $\textit{U(z)}$ of the a-type structure with $\textit{F}_{appl}$ from
\textit{6.0}\textrm{ V/$\mu$m} to \textit{15.0}\textrm{ V/$\mu$m}, by step of \textit{1.0}\textrm{ V/$\mu$m}. The
origin is at one of the last nucleus of the SWCNT. The lower barrier is corresponding to larger $\textit{F}_{appl}$.
(b) $\textit{U(z)}$ of the three ending structures in the applied field of \textit{7.0}\textrm{ V/$\mu$m}. The solid,
dashed, and dotted curves are corresponding to the a-type, b-type, and c-type structures, respectively. Here the origin
is on the surface of the substrate.}
\end{figure}

The superposition of Coulomb potential of all charges in the tube and their images in the substrate, together with the
potential of \textit{F}$_{appl}$, determines the AVB for the electron emission. In FIG. \ref{fig_5}(a), we plot the
electrostatic potential \textit{U(z)} of the a-type structure for various \textit{F}$_{appl}$. The \textit{Z} axis has
its origin at the last atom of the SWCNT and is parallel to the direction of tube axis. The dependence of AVB on
\textit{F}$_{appl}$ is remarkable. The barrier potential \textit{U(z)} of the three ending structures are compared in
FIG. \ref{fig_5}(b) for \textit{7.0}$\textrm{ V/$\mu$m}$. Comparing to the AVB of the a-type structure, the AVB of the
b-type structure is lower but thicker. This partly results from the difference of the apex dipoles (see TABLE V).
Suppose that the applied field would not change significantly the intrinsic electronic structure of the SWCNT. Then in
the a-type (b-type) structure, the atoms of the first layer have negative (positive) dipoles, and the atoms of the
second layer have positive (negative) dipoles. The potential close to the apex is mostly affected by the atoms of the
first layer and is enhanced (suppressed). Apart from the apex in some distance (7 angstroms), the dipoles of the first
layer would be screened by those of the second layer, the potential of the a-type structure could not be distinguished
from that of the b-type. For the c-type structure, the dipoles of the first two layers are negative, the net negative
dipole and the large negative excess charge at the apex make the AVB high and thick.

\begin{figure}
\includegraphics[scale=0.8]{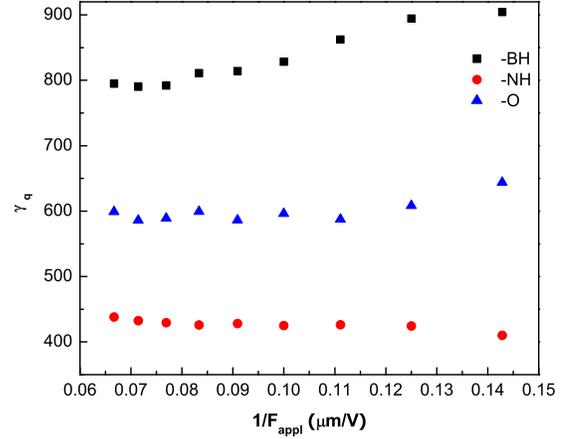}
\caption{\label{fig_6} The field enhancement factor versus the inverse of applied field. The squares, circles, and
triangles are data of the simulations of the a-type, b-type, and c-type structures, respectively.}
\end{figure}

The field enhancement factor as a feature of the AVB is highly concerned in the FE applications. For the SWCNT,
however, it is difficult to define a quantity that corresponds exactly to the field enhancement factor as usually
defined for planar metallic emitters. Denote the maximum local field by \textit{F}$_{apex}$. Here the field enhancement
factor $\gammaup_{q}$ is defined as \textit{F}$_{apex}$/\textit{F}$_{appl}$. The variation of this field enhancement
factor with the value of 1/\textit{F}$_{appl}$ is shown in FIG. \ref{fig_6}. Note that $\gammaup_q$ of these structures
do not change significantly with applied fields, but are obviously different from each other. It reflects the important
influence of the adsorbates on FE.

\section{\label{sec:level}Emission currents and emission paths}

\begin{figure}
\includegraphics[scale=0.8]{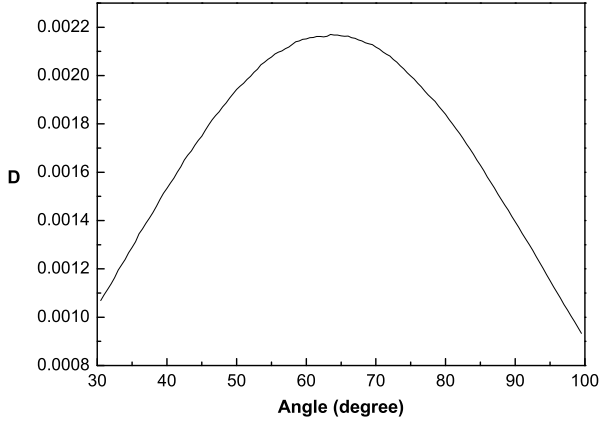}
\caption{\label{fig_7} \textit{D} versus path angle for the a-type structure in \textit{12.0}\textrm{ V/$\mu$m}.}
\end{figure}

\begin{figure}
\includegraphics[scale=0.8]{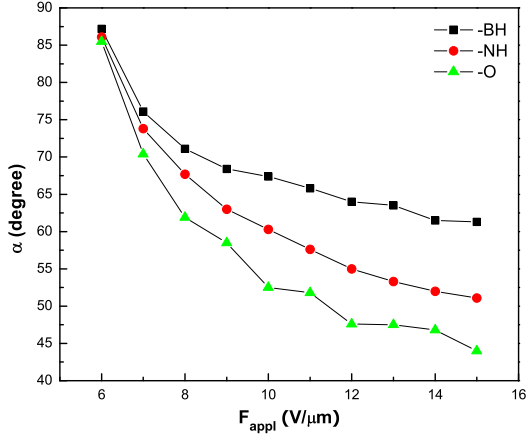}
\caption{\label{fig_8} The angle of the most probable path versus the applied field. The squares, circles, and
triangles are the data corresponding to the a-type, b-type, and c-type structures respectively.}
\end{figure}

According to the quantum mechanics, electrons may emit to the vacuum along any path with non-zero probability. In the
semi-classical approximation, only the path of the least (maximum) action is considered. This path will be referred to
as the most probable path (MPP). Since the a-type and b-type ending structures have the thinnest AVB in front of the
hydrogen atoms, MPP of these structures should start from the hydrogen atom. The thinnest AVB of the c-type structure
starts from the top carbon atoms and goes outward along a path through the middle point of two adjacent oxygen atoms.
For simplicity, we assumed that the electrons go through the barrier in straight lines. The path angles
({$\mathit{\alpha}$}) of three ending structures are presented schematically in FIG. \ref{fig_2}. Along the MPP, the
transmission coefficient (\textit{D}) can be estimated by the WKB approximation

\begin{equation}\label{eq_2}
D = \exp\Bigl[-\frac{2}{\hbar}\int\sqrt{2mV(z)}\mathrm{d}z\Bigr],
\end{equation}

\noindent where \textit{V(z)} is the electron energy potential related to the Fermi level, and the integral is over the
classical forbidden region, i.e., where $V(z)>0$.  We has assumed that the electrons possess the Fermi level. If the
path is not the MPP, the value of \textit{D} by Eq. \ref{eq_2} should be smaller. For instance,  In FIG. \ref{fig_7},
\textit{D} is plotted against the path angle for the a-type structure in \textit{12.0}\textrm{ V/$\mu$m}. The angle of
the MPP is determined by the maximum of \textit{D}. Since the AVB is field-dependent, the MPP would change its
orientation with the applied fields. The angles of the MPP versus the applied fields are plotted in FIG. \ref{fig_8}.
The squares, circles, and triangles correspond to the a-type, b-type, and c-type structures respectively. One sees that
the angle of the MPP decreases as the applied field increases. This phenomenon would be related to the fact that higher
applied field induces more excess electrons at the apex.

With \textit{D} in hand, the emission current is estimated by

\begin{equation}\label{eq:barwq}
I = \nu q_{exc}D,
\end{equation}

\noindent where $q_{exc}$ are the excess electrons around the atoms from which the MPPs start, and $\nu$ is the
collision frequency that can be estimated from the average kinetic energy of $\pi^*$ electrons as $E_k(\pi^*) / h$.
Another way to estimate the collision frequency is to use the uncertainty relation

\begin{equation}\label{eq:barwq}
\nu = \frac{E_k}{h} = \frac{h}{32\pi^2 m<\bigtriangleup r^2>},
\end{equation}

\noindent where $<\bigtriangleup r^2>$ is the uncertainty of the radial coordinate. Its numerical value can be
estimated from the density of excess electrons (FIG. \ref{fig_3}). The collision frequency estimated by two methods has
the same order of $\mathit{10^{14}}$\textrm{ Hz}.

\begin{figure}
\includegraphics[scale=0.8]{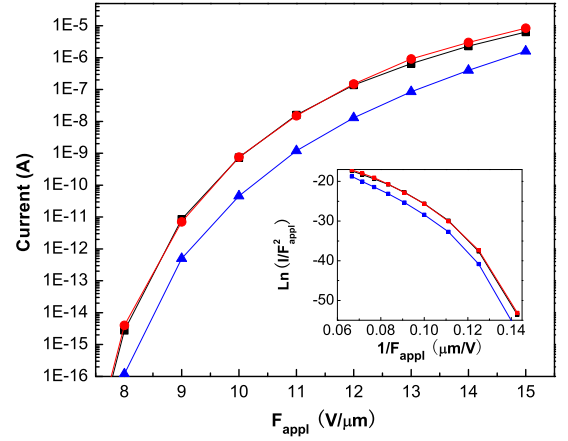}
\caption{\label{fig_9} The currents versus applied field. The squares, the circles, and the triangles are the currents
corresponding to the a-type, b-type, and c-type structures, respectively. The inset is the Fowler-Nordheim plot where
1/\textit{$F_{appl}$} is the horizontal axis and log(I/\textit{$F^2_{appl}$}) the vertical axis.}
\end{figure}

The diagram of the emission current versus applied field is presented in FIG. \ref{fig_9}, where squares, circles, and
triangles are for the a-type, b-type, and c-type structures, respectively. For each structure, the emission current is
the summation of the currents along five MPPs, each MPP is as defined in FIG. \ref{fig_2}.

To confirm that the emission is dominated by the MPPs, we have also calculated the emission current along the path
starting from a charge center in between two nearby MPPs. The results for \textit{$F_{appl}$ = }\textit{12.0}\textrm{
V/$\mu$m} are compared in TABLE VI. The contribution of the MPP is one order or so larger than that of the path not a
MPP.

\begin{table}
\caption{\label{tab_VI} The emission currents (A) from the MPP and from the path in between two MPPs.}
\newcommand{\rb}[1]{\raisebox{-2.0ex}[0pt]{#1}}
\newcommand{\rc}[1]{\raisebox{1.0ex}[0pt]{#1}}
\begin{ruledtabular}
\begin{tabular}{ccccc}
\multicolumn{2}{c}{Structures} & -BH & -NH & -O \\
\hline \rb{current} & MPP & 1.4E-7 & 1.5E-7 & 1.3E-8 \\
\cline{2-5} & path not a MPP & 8.7E-9 & 1.4E-8 & 2.1E-9 \\
\end{tabular}
\end{ruledtabular}
\end{table}

\section{\label{sec:level}Conclusions and Discussions}

A one micrometer long (5, 5) single-walled carbon nanotube (SWCNT) has been simulated by the multi-scale method, with
the interest focused at the effects of ending structures. The apex-vacuum barriers of the SWCNTs ended by -BH, -NH, and
-O respectively have been obtained. The local field enhancement factor is different from one ending structure to
another; all are much smaller than the prediction of the classical model. The lowering of the apex-vacuum barrier by
the applied field, besides the local field enhancement, as the essential mechanism is confirmed to be responsible for
the low turn-on field of SWCNTs.

According to our simulations, the SWCNTs terminated with diatoms whose outer atoms have smaller electronegativity would
be superior in field emission and have more stable in structure. The turn-on field (~\textit{11.5}\textrm{ V/$\mu$m})
of the SWCNT terminated with -BH or -NH is smaller than that terminated by oxygen. In the same applied field, the
current of oxygen terminated SWCNT is an order or more weaker than those terminated by -BH and -NH. The simulation
suggests that the binding energy of the diatom decorated SWCNTs is about \textit{4.5} times the binding energy of the
oxygen saturated SWCNT and more than double of the hydrogen saturated one. The electronegativity of the adsorbates may
partly explain the feature of the apex-vacuum barrier and give a general hint for the structure optimization. The
orientation of the most probable emission path depends on the applied field significantly. This observation would
provide a mean for extracting the atomic features of the apex.

One would have noticed that the best turn-on field we estimated for the -BH ending structure is still too large than
most experimental results that is about \textit{1}-to-\textit{5}\textrm{ V/$\mu$m}. \cite{a32} There would be a number
of reasons, besides the uncontrollable error of the simulation. First, the emission current could be larger if one
takes into account the emission from the paths other than the most probable paths we have considered. Secondly, in our
calculation, we have ignored the exchange and correlation effect (i.e., the image force) on the emitting electron. The
prediction would be improved by including this effect. However, to our knowledge, it is still difficult to include the
exchange and correlation effect in a quantum mechanical simulation of emission current of SWCNT. \cite{a33} Last but
not least, most, if not all, CNTs used in the FE experiments are multi-walled CNTs and the chirality is not
controllable. The experimental observed emission current would be higher since the multi-walled CNTs would have larger
field enhancement factor as it would resemble more a classical metal rod. The above argument implies the existence of
discrepancy between our simulational result and the experimental observations. With optimism, our simulation would have
revealed some virtual elements involved in the field emission of SWCNT with the atomic decoration, at least
qualitatively.
\newline

\textbf{Acknowledgement.} The authors thank Chris J. Edgcombe and Richard G. Forbes for the valuable discussions; and
gratefully acknowledge the support from the National Natural Science Foundation of China (the Distinguished Creative
Group Project; Grant No. 10674182, 90103028, 90306016) and  from Hong Kong Research Grant Council (HKU 7010/03P, HKU
7012/04P).

\end{document}